\documentclass[aps,pre,twocolumn,showpacs,groupedaddress]{revtex4}

\usepackage{subfigure}
\usepackage{graphicx}
\usepackage{amssymb}
\usepackage{wasysym}

\begin{document}

\title{Dense bubble flow in a silo: an unusual flow of a dispersed medium}
\author{Yann Bertho, Christophe Becco, and Nicolas Vandewalle}
\affiliation{Group for Research and Applications in Statistical
Physics (GRASP), Institut de Physique B5a, Universit\'e de
Li\`ege, 4000 Li\`ege, BELGIUM}

\begin{abstract}
The dense flow of air bubbles in a two-dimensional silo (through
an aperture $D$) filled with a liquid is studied experimentally. A
particle tracking technique has been used to bring out the main
properties of the flow: displacements of the bubbles, transverse
and axial velocities. The behavior of the air bubbles is observed
to present similarities with non-deformable solid grains in a
granular flow. Nevertheless, a correlation between the bubble
velocities and their deformations has been evidenced. Moreover, a
new discharge law (Beverloo-like) must be considered for such a
system, where the flow rate is observed to vary as $D^{1/2}$ and
depends on the deformability of the particles.

\end{abstract}

\pacs{47.50.+d, 47.60.+i, 82.70.-y, 83.50.-v, 83.80.-k}

\maketitle

Structured fluids like granular materials or aqueous foams
received much attention from physicists during the last decade.
Indeed, they are ubiquitous in nature and in industrial processes
\cite{Duran00b, Weaire99}, and exhibit striking rheological
behaviors as compared to conventional liquids. These materials
involve multiple physical processes and cooperative phenomena such
as shear banding, particle interactions and the formation of
arches redirecting the forces inside the material \cite{Janssen95,
Bertho03}. Such processes are then able to provoke intermittent
flows \cite{Bertho03b} or jamming \cite{Rodts05} because of the
topological constraints that develop since the particles (grains
or bubbles) press against each other. This contributes to the
complex flow behavior of granular media and foams.

\begin{figure}[b!]
\begin{center}
\includegraphics[width=8.6cm]{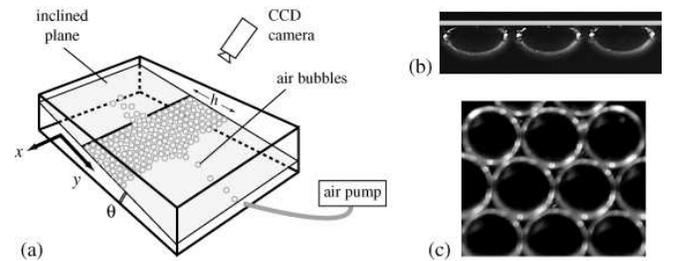}
\caption{(a)~Sketch of the experimental setup. A transparent
inclined plane is immersed into water and tilted at an angle
$\theta$. Small air bubbles are injected from below at the bottom
of the plane and rise towards the top where they aggregate.
Motions of the bubbles are recorded by a CCD camera. (b)~Side
view: the bubbles present an oblate shape. (c)~Top view: the
bubbles have a circular shape in the bulk.} \label{Fig01}
\end{center}
\end{figure}

The present study deals with bubble assemblies where the bubbles
are predominantly oblate so that the contacts between two adjacent
bubbles are reduced (see Fig.~\ref{Fig01}b). The liquid fraction
in the material is high enough to consider the diffusion of air
through the bubble boundaries as negligible at the time scale of
the experiments.

Bubble assemblies can also be seen as a granular medium where
contacting particles are characterized by a nearly zero friction.
Thus, we have a remarkable system with a low energy dissipation at
the contacts in comparison with traditional grain assemblies. Both
granular media and foams are model systems for divided materials.
Nevertheless, even if bubble flows present qualitative analogies
with granular flows, we must keep in mind that the nature of the
interactions between the particles is totally different. Moreover,
in contrast with granular materials, the bubbles may undergo
strong deformations when they are subjected to a constraint. Many
works deal with the study of flowing foams (velocity profiles,
bubble deformations) in a Couette geometry \cite{Lauridsen04,
Janiaud05} or through a constriction \cite{Asipauskas03}. In
contrast with this latter paper dealing with a dry foam flowing in
a Hele-Shaw cell, our interest concerns wet foams flowing between
a plate and the liquid surface. Instantaneous measurements of
velocity and deformation of the bubbles are reported. The
dependence of the bubble flow rate regarding the aperture or the
mean bubble velocity is analyzed through the comparison with
granular materials in a same geometry.

In this paper, the flow of a two-dimensional (2D) foam is analyzed
experimentally \cite{Bragg47, Vandewalle04}. The setup consists of
a transparent inclined glass plane of width $W=130$\,mm which is
immersed into water (Fig.~\ref{Fig01}a). The tilt angle $\theta$
of the plane can be adjusted by a fine screwing system.
Quasi-monodisperse air bubbles are then injected at the bottom of
the inclined plane by blowing air through a needle (the height $h$
of the piling is kept roughly constant during the experiments
$h\simeq$ 200\,mm). Note that the tilt angle is small ($\theta$
ranges from 0$^\circ$ to 1$^\circ$) to ensure that only one layer
of bubbles is created in the perpendicular direction to the plane
and to decrease the influence of the gravity (\emph{i.e.} liquid
drainage). Bubble size is controlled by an air pump and is kept
roughly constant in the present study [diameter $d=(5.4 \pm
0.3)$\,mm]. In order to avoid the coalescence of the bubbles,
dish-washing liquid based on anionic surfactant AEOS-2EO is mixed
into water \cite{Broze99}. The surface tension of the solution is
$\gamma\simeq 25$\,mN\,m$^{-1}$ leading to a capillary length
$\kappa^{-1}=\sqrt{\gamma/(\rho g)}\simeq 1.6$\,mm. In our
experiment, the bubble are larger than $\kappa^{-1}$ and
consequently adopt an oblate shape (Fig.~\ref{Fig01}b) so that the
contact between two adjacent bubbles is reduced. Due to buoyancy,
the bubbles rise underneath the inclined plane and tend to pack on
a transverse wall placed at the top of the plane. An orifice of
width $D$ at the center ($x_0$, $y_0$) of this obstacle allows the
bubbles to empty out the silo. Our experiment is then analogous to
the gravity-driven granular flow in a flat bottomed 2D silo,
emptying out of a central orifice \cite{Medina98}. Top views of
the bubble packing are recorded through the transparent tilted
plane by means of a CCD camera at a frame rate of 50 frames per
second. In order to quantify and to measure the main properties of
the flow, the motion of each bubble ($\approx$ 2000 bubbles per
frame) has been tracked through image analysis.\ \\
\begin{figure}[t]
\begin{center}
\includegraphics[width=8.6cm]{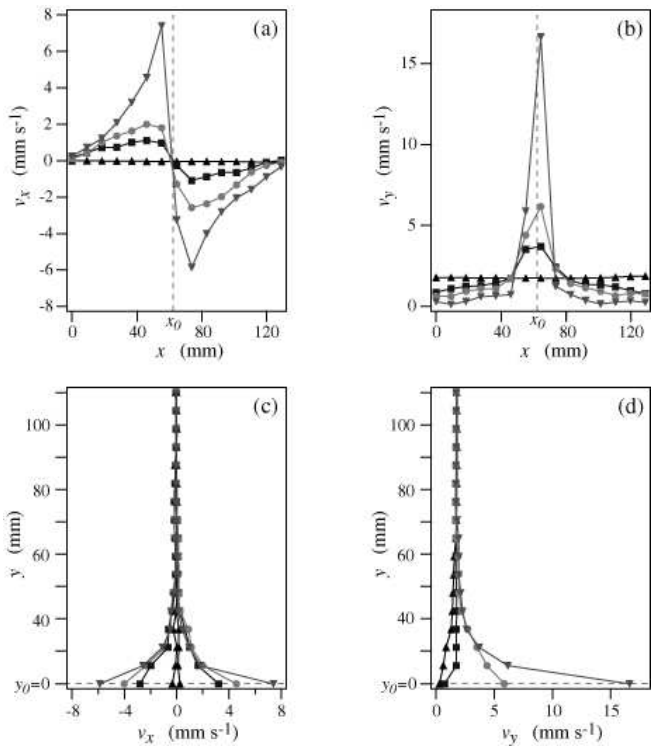}
\caption{Transverse $v_x$ and longitudinal $v_y$ velocity profiles
during the discharge of a silo ($\theta\simeq 0.63^\circ , D\simeq
8.6$\,mm). (--~--) position ($x_0$, $y_0$) of the center of the
outlet of the silo. The different curves ($\blacktriangledown$,
$\bullet$, $\blacksquare$, $\blacktriangle$) correspond
respectively to velocity profiles at increasing distances from
$y_0=0$ (Figs.~\ref{Fig02}a and \ref{Fig02}b) and increasing
distances from $x_0$ (Figs.~\ref{Fig02}c and \ref{Fig02}d).}
\label{Fig02}
\end{center}
\end{figure}

Figure~\ref{Fig02} displays typical transverse $v_x$ and
longitudinal $v_y$ velocities of the bubbles as a function of $x$
and $y$ at different heights in the packing. In the top part of
the silo [($\blacktriangle$) in Fig.~\ref{Fig02}b], a block motion
is observed: the longitudinal velocity $v_y$ remains constant
reflecting low interactions between the bubbles and the side walls
(slipping motion). As $y$ decreases (\emph{i.e.} the bubbles move
towards the outlet), the flow takes place in a triangular-shaped
domain where the fastest bubbles are located in the center of the
silo and the slowest or motionless ones at the walls
[($\blacktriangledown$) in Fig.~\ref{Fig02}b]. This is in
agreement with the `V-shape' of mobile grains observed during the
flow of granular materials in silos or hourglasses
\cite{Hirshfeld97,Medina98}. Far from the outlet ($y\gtrsim 15d$),
bubbles have a nearly zero transverse velocity (Fig.~\ref{Fig02}c)
and a constant longitudinal velocity $v_y$ (Fig.~\ref{Fig02}d).
For lower values of $y$, the velocity distribution of the bubbles
deep inside the silo is totally modified: a transverse component
of velocity $v_x$ appears in the flow conducting the bubbles
towards the orifice while $v_y$ increases strongly near the
outlet.

\begin{figure}[t!]
\begin{center}
\includegraphics[width=7cm]{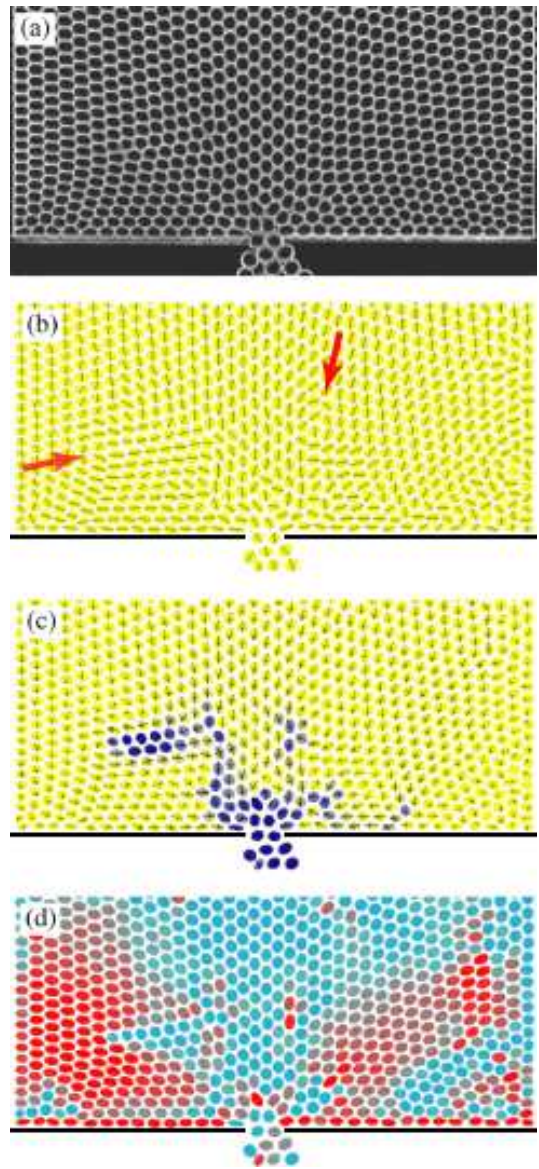}
\caption{(Color online) Typical recordings of the bubble
discharge: (a)~Snapshot of the experimental flow. (b)~Displacement
field. The arrows point out a dislocation and a recirculation
zone. (c)~Velocity field. The fastest bubbles appear in blue (or
darker). (d)~Deformation field. The more constrained bubbles
appear in red (or darker) while circular bubbles appear in blue.}
\label{Fig03}
\end{center}
\end{figure}
A zoom of the bottom part of the silo of height 60\,mm is shown in
Fig.~\ref{Fig03}a. Image analysis allows one to extract the
position of each bubble in the stack and evaluate the displacement
profile (Fig.~\ref{Fig03}b) and the velocity field
(Fig.~\ref{Fig03}c) during the flow. Figure~\ref{Fig03}d displays
the deformation field and provides information concerning the
constraints undergone inside the dense bubble assembly. In
contrast with granular materials where the flow occurs only in a
cone-shaped central region of the silo (while the grains located
in the regions near the side walls are motionless)
\cite{Hirshfeld97,Medina98}, note that motions of bubbles are
detected everywhere in the silo. Moreover, the bubble tracking put
into evidence recirculation zones and the propagation of
dislocations (shear bands) during the bubble flow. As expected in
two-dimensional nearly monodisperse flows, bubbles tend to pack in
an ordered hexagonal structure at the top of the bubble packing
(Fig.~\ref{Fig03}a); these organized domains are separated by the
dislocations. Motions of blocks of bubbles are observed and look
like crystal domains. Those blocks are sliced by fast-moving
defects along bubble lines. The movement is not simultaneous along
the whole row but begins at one end with the appearance of a
dislocation running along the slip line. This process is initiated
at the orifice of the silo and propagates towards the bulk after
many rebounds on the side walls. Both recirculation zone and
dislocation appear clearly as pointed out by the arrows in
Fig.~\ref{Fig03}b.

The instantaneous velocity $v=(v_x^2+v_y^2)^{1/2}$ of each bubble
has been computed and superimposed on the images of the flow (the
darkest -- or blue -- bubbles corresponding to the fastest ones).
The recirculation regions and the dislocation observed in
Fig.~\ref{Fig03}b are characterized by a mean velocity higher than
in the rest of the silo (Fig.~\ref{Fig03}c).

The deformation of a bubble is evaluated by fitting the bubble
shape by an ellipse and computing its eccentricity \cite{Ybert02}.
As pointed out in Fig.~\ref{Fig03}d, the fastest bubbles are
observed to correspond to the less deformed ones. Moving bubbles
press against surrounding ones and deform them. This is especially
the case near the dislocation. Therefore, in such dense bubble
flows, spherical bubbles are located far from the outlet and
propagate in a block motion at a constant velocity or in the
dislocation, while flattest ones correspond to crushed bubbles
against the bottom wall or constraint bubbles inside the pile.
\ \\
\begin{figure}[t!]
\begin{center}
\includegraphics[width=8.6cm]{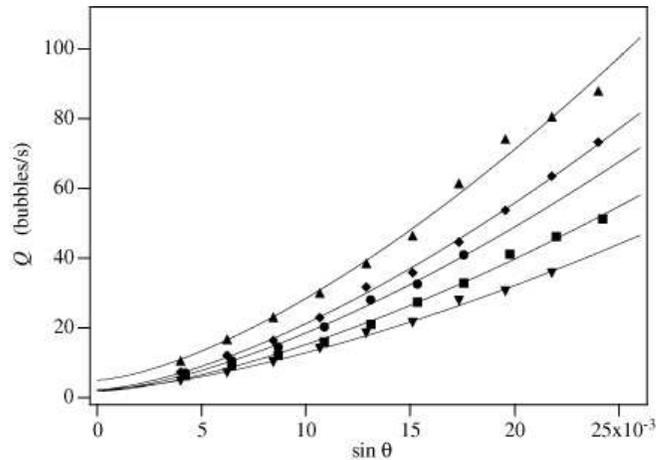}
\caption{Bubble flow rate $Q$ as a function of the tilt angle
$\theta$ of the plane. The curves correspond to different
apertures $D$ of the silo: ($\blacktriangledown$)~$D=6.3$\,mm,
($\blacksquare$)~$D=8.0$\,mm, ($\bullet$)~$D=9.7$\,mm,
($\blacklozenge$)~$D=12.8$\,mm, ($\blacktriangle$)~$D=19.9$\,mm;
(---)~best fit of the experimental data with $Q\propto (\sin
\theta)^{3/2}$.} \label{Fig04}
\end{center}
\end{figure}

A classic paper on the flow of particles through orifices is that
of Beverloo \emph{et al.} \cite{Beverloo61}. Using a
straightforward dimensional analysis, they pointed out that the
grain flow rate $Q_g$ is a power law in $D$ (where $D$ is the
diameter of the orifice). The power depends on the dimensionality
of the flow \cite{Hirshfeld97, Beverloo61, Mills96}, so that for a
2D flow $Q_g\propto g^{1/2}D^{3/2}$. A corrective term is usually
added to take into account the mean diameter of the grains $d_g$:
\begin{equation}
Q_g\propto g^{1/2}(D-kd_g)^{3/2},\label{Beverloo}
\end{equation}
where $k$ is an empirical constant depending on the particle shape
and the grain interactions ($k\simeq 1.5$ for beads). This law
characterizing the dynamical properties of the grain flow does not
invoke explicitly dissipative mechanisms. However, we expect that
$k$ and the value of the exponent 3/2 contain implicitly friction
and deformation of the grains even if the dependence of these
parameters with friction and deformability is still an open
question.

The bubble flow rate $Q$ has been evaluated by counting the number
of bubbles flowing out of the silo, at different tilt angles
$\theta$ of the plane. Figure~\ref{Fig04} shows that $Q$ varies as
$(\sin \theta)^{3/2}$, where the tilt angle plays a similar part
to gravity encountered in the traditional Beverloo's law. The
behavior of bubble flow through an aperture is thus drastically
different from the one of a granular flow for which $Q_g$ varies
as $g^{1/2}$. This is probably due to the nature of forces between
the particles and with the walls: the frictions inside a granular
medium are much more important than in a bubble flow and
consequently slow down the silo discharge.

\begin{figure}[t]
\begin{center}
\includegraphics[width=8.6cm]{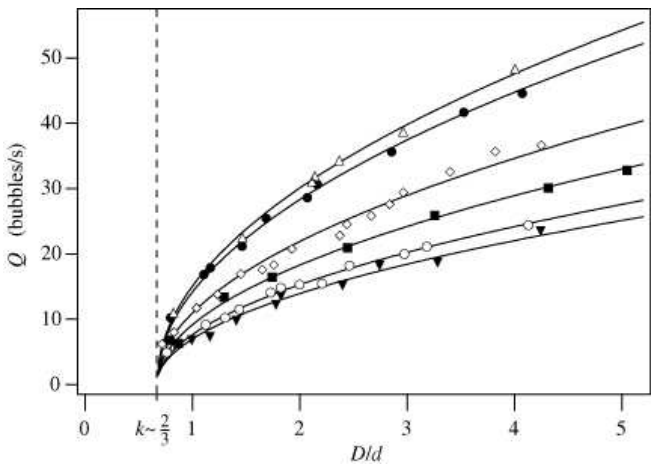}
\caption{Bubble flow rate $Q$ as a function of the aperture $D$ of
the silo (normalized by the mean bead diameter $d$). The curves
correspond to different tilt angles $\theta$ of the plane:
($\blacktriangledown$)~$\theta=0.46^\circ$,
($\circ$)~$\theta=0.50^\circ$,
($\blacksquare$)~$\theta=0.56^\circ$,
($\diamond$)~$\theta=0.63^\circ$, ($\bullet$)~$\theta=0.75^\circ$,
($\vartriangle$)~$\theta=0.82^\circ$; (---)~best fit of the
experimental data with $Q\propto (D/d - k)^{1/2}$; $k$ presents
very low dispersion and verifies $k=(0.66\pm 0.03)\simeq
\frac{2}{3}$.} \label{Fig05}
\end{center}
\end{figure}

Figure~\ref{Fig05} displays the bubble flow rate $Q$ as a function
of the normalized aperture of the silo $D/d$. In contrast with
granular materials for which the flow rate $Q_g$ vanishes for
$D/d\lesssim 1.5$ due to the formation of arches
[Eq.~(\ref{Beverloo})], $Q$ is observed to be non-zero for smaller
values of the aperture. Bubbles can pass through an orifice of the
order of $2d/3$ after having undergone strong shape deformations.
The value of $k\simeq 2/3$ should then depend mainly on the bubble
size and the surface tension of the liquid; these assumptions will
receive special care in a subsequent work. Moreover, note that $Q$
varies as $(D/d)^{1/2}$ which is significantly different from the
case of a non-deformable 2D granular material flowing down a silo
where the power is 3/2.

The bubbles at the outlet of the silo are affected by the buoyancy
$B$ coming from the bubbles located behind them. This buoyancy
presents a non-trivial law because it does not result directly
from the sum of each individual bubble in the stack: a part of the
energy is dissipated through the deformation of the bubbles. Let
assume that the driving force is the buoyancy component along the
plane, so that $B\propto g\sin \theta$.

The dissipation relationship linking the drag force $f$ to a given
velocity $v$ is non-trivial, and known to involve rather subtle
hydrodynamics \cite{Bretherton61, Denkov05}. If Archimedes' forces
were balanced by a Stokes' viscous force ($f\propto \eta dv$),
this would provide a constant rise velocity proportional to $\sin
\theta$. But for a single bubble creeping along a slightly
inclined plane immersed in a viscous fluid, it has been shown that
the creeping velocity $v$ does not vary linearly with the tilt
angle $\sin \theta$ \cite{Aussillous02}. Moreover, the viscous
dissipation is known to be sensitive to the surface mobility of
the surfactant \cite{Denkov05}. For our surfactant which leads to
tangentially mobile bubble surfaces, if we consider the viscous
dissipation associated with the formation of a lubricating film
between the single bubble and the plane, the viscous force $f$ is
observed to vary as $\gamma^{1/12}(\eta v)^{2/3}$.

Assuming in a first approximation that the dynamics of the flow is
mainly governed by these two forces (neglecting therefore the
interactions between the bubbles), the equilibrium between the
buoyancy and the drag force leads to a terminal velocity $v$
proportional to $(\sin \theta)^{3/2}$. Therefore, the bubble flow
rate $Q=vD/d^2$ can be written
\begin{equation}
Q\propto \frac{(g\sin \theta)^{3/2}}{\eta \gamma ^{1/8}},
\label{eqQ2}
\end{equation}
which agrees with behaviors emphasized in Fig.~\ref{Fig04}.
Moreover, as expected, the bubble flow rate is then observed to
decrease when the viscosity $\eta$ of the fluid is increased. This
straightforward model reproduces the main effects observed
experimentally and enhances the fact that a new Beverloo's law
must be considered in the particular case of bubble flows and more
generally the flow of deformable dispersed media.
\begin{figure}[t]
\begin{center}
\includegraphics[width=8.6cm]{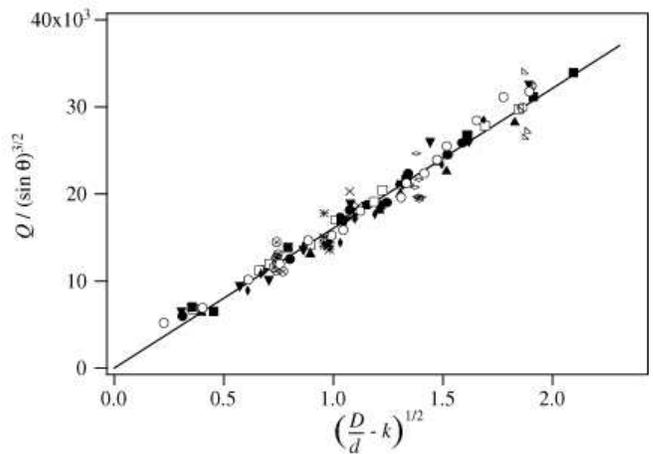}
\caption{Rescaling of all measurements using Eq.~(\ref{eqQ3}) for
the data presented in Figs.~\ref{Fig04} and \ref{Fig05}.}
\label{Fig06}
\end{center}
\end{figure}

Figure~\ref{Fig06} displays a rescaling of all measurements and
confirms that the Beverloo's law for a deformable dispersed medium
is given by:
\begin{equation}
Q\propto (g\sin\theta)^{3/2}\left (\frac{D}{d}-k \right
)^{1/2},\label{eqQ3}
\end{equation}
where $k<1$ is related to the deformation ability of the particles
and the friction between particles. More experiments must now be
performed to distinguish the influence of both friction and
deformation on $k$, and on the value of the exponent. Numerical
simulations are currently performed on this topics and will be the
object of a subsequent publication. Finally, following the
approach proposed by Asipauskas \emph{et al.} \cite{Asipauskas03},
the study of the influence of the bubble deformation on the flux
measured at the outlet of the silo would provide a good complement
to the present study.

In summary, the present experiment shows that the velocity
profiles for bubble flows in silos present qualitative
similarities with a granular flow but with sliding motions at the
walls. A correlation between the velocity and the deformation of
the bubbles has been observed. Moreover, we demonstrate that a new
Beverloo's law must be considered for the specific case of
deformable dispersed media flowing out a silo, where the flow rate
varies as $D^{1/2}$ and depends on the nature of the interactions
between particles and their capability to deform themselves when
they are submitted to a constraint.

The authors thank G. Broze and S. Dorbolo for helpful discussions
during this work. Y. B. benefits from a postdoctoral fellowship
supported by the ARC programme number 02/07-293 and ESA MAP
``AO-99-108".

\end{document}